# Ultrafast quantum key distribution using fully parallelized quantum channels

**ROBIN TERHAAR**[1,5,6], **JASPER RÖDIGER**[2,5], **MATTHIAS HÄUßLER**[1,5], **MICHAEL WAHL**[3,5], **HELGE GEHRING**[1], **MARTIN A. WOLFF**[1], **FABIAN BEUTEL**[1], **WLADICK HARTMANN**[1], **NICOLAI WALTER**[1], **JONAS HANKE**[2], **PETER HANNE**[2], **NINO WALENTA**[2], **MAXIMILIAN DIEDRICH**[2], **NICOLAS PERLOT**[2], **MAX TILLMANN**[3], **TINO RÖHLICKE**[3], **MAHDI AHANGARIANABHARI**[3], **CARSTEN SCHUCK**[1] AND **WOLFRAM H.P. PERNICE**[4,7]

[1]*Institute of Physics, University of Münster, Heisenbergstraße 11, 48149 Münster, Germany*
[2]*Fraunhofer Heinrich Hertz Institute, Einsteinufer 37, 10587 Berlin, Germany*
[3]*PicoQuant GmbH, Rudower Chaussee 29, 12489 Berlin, Germany*
[4]*Universität Heidelberg, Kirchhoff-Institut für Physik, Im Neuenheimer Feld 227, 69120 Heidelberg, Germany*
[5]*These authors contributed equally*
[6]*robin.terhaar@uni-muenster.de*
[7]*wolfram.pernice@kip.uni-heidelberg.de*

**Abstract:** The field of quantum information processing offers secure communication protected by the laws of quantum mechanics and is on the verge of finding wider application for information transfer of sensitive data. To overcome the obstacle of inadequate cost-efficiency, extensive research is being done on the many components required for high data throughput using quantum key distribution (QKD). Aiming for an application-oriented solution, we report on the realization of a multichannel QKD system for plug-and-play high-bandwidth secure communication at telecom wavelength. For this purpose, a rack-sized multichannel superconducting nanowire single photon detector (SNSPD) system, as well as a highly parallelized time-correlated single photon counting (TCSPC) unit have been developed and linked to an FPGA-controlled QKD evaluation setup allowing for continuous operation and achieving high secret key rates using a coherent-one-way protocol.

## 1. Introduction

As secure communication becomes increasingly important for today's society, encryption technology becomes more and more omnipresent. Especially recent advances in the field of quantum computing put advanced classical encryption approaches such as the RSA public-key cryptosystem at risk [1,2]. Therefore, it is only logical that the field of quantum cryptography has become the fastest-growing area in information science, bringing up many different candidates for quantum key distribution (QKD) protocols [3,4]. Depending on the used QKD protocols, the required components need to fulfill different criteria to be applicable, but the common bottleneck for all QKD implementations is the imperfections of the single photon detector [5]. Scalability, cost-efficiency, and overall performance for single photon detectors have been difficult to reconcile ever since. The superconducting nanowire single photon detector (SNSPD), achieving near unity efficiency [6,7], up to GHz count rates [8] and single digit picosecond jitter [9], excellently meets the necessities for the field of quantum information [10], especially for applications using longer wavelengths in the near-infrared (NIR) that are used for optical communication networks.

Independently from the detector technology, capable readout electronics are required to be able to scale up in performance or bandwidth respectively, particularly if time-bin encoding is used on multiple channels. With modern time-correlated single photon counting (TCSPC) units based on time-to-digital converters (TDCs), dead times as short as 650 ps with picosecond timing resolution can be achieved for multiple inputs [11]. While TCSPC systems are typically used for basic fluorescence lifetime measurements that employ hardware histogramming for data reduction, time-bin encoding requires to have live access to the individual arrival time of each detected photon. To handle the resulting digital data rates, such systems must employ advanced data bus technologies to stream the tag data in real-time to a host PC.

As outlined earlier, one field, which will profit from highly parallelized single photon detection, is the field of QKD. In combination with the one-time-pad (OTP), provable information-theoretic secure communication becomes possible [12]. Especially, a high number of detectors can improve performance tremendously, e.g., utilizing multiplexing. Like classical communication, multiplexing can be used to increase the capacity per optical link.

Making use of the inherent scalability of advanced SNSPDs and modern TCSPC electronics towards high bandwidth QKD, we designed a highly parallel architecture for multiplexing with up to 64 channels. To handle the resulting data rates in real-time, we furthermore designed a new generation of time tagging electronics that can deliver their data streams not only to a host PC but also to an external FPGA through high-speed serial links capable of handling a total photon rate of up to 1.6 GHz.

This SNSPD and TCSPC technology was then utilized to implement QKD, namely the coherent one-way (COW) QKD protocol. The COW protocol, which was introduced by Gisin et al. in 2004 [12], belongs to the distributed phase reference protocols and thus is based on time-phase encoding [13]. To benefit from a large number of detectors for increasing the performance, multiple QKD channels were multiplexed with wavelength division multiplexing (WDM). In a straightforward implementation, each QKD receiver would contain an interferometer, respectively, to measure in the phase basis. To decrease the complexity and cost of the QKD system, a simplified receiver design was developed and implemented, where only one interferometer is needed on the receiver side without the need for additional phase correction. Each QKD channel uses two SNSPD channels, and the evaluation of a QKD transmission is performed in real-time on the external FPGA.

## 2. Multichannel Implementations for QKD

### 2.1 *COW-protocol with WDM for highly parallelized QKD*

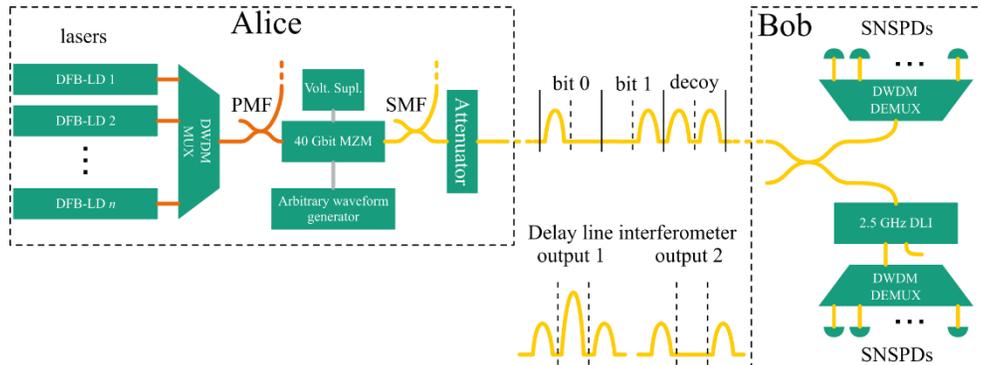

Figure 1: Schematic arrangement of the QKD setup components with the QKD sender unit named Alice and the QKD receiver unit named Bob. The QKD sender employs a wavelength multiplexing scheme, using multiple DFB lasers as light sources that are combined in a polarization maintaining (PMF) DWDM unit. The combined signals are modified by an optical intensity modulator (Oclaro SD40), and the total intensity is damped down to single-photon level using an optical attenuator. The receiver unit Bob splits the signal to measure time and phase basis and divides the different channels according to ITU before they are detected using SNSPDs.

In the COW protocol, the sender ("Alice") comprises a laser and an intensity modulator for creating the three different symbols in the time domain, with each symbol consisting of two time-bins. These three symbols are the "0" and "1" symbols, where one pulse is in the early

(respectively late) time-bin, and the "decoy" symbol, where both time-bins contain a pulse (see Figure 1). Before transmitting the stream of symbols to the receiver ("Bob"), they are attenuated below single-photon level. At Bob's setup, the symbols are measured either in the time basis or in the phase basis by triggering on the SNSPD's rising edge with the multichannel TCSPC unit. In the time basis, the arrival time of the incoming photons is measured to distinguish between "0" and "1" symbols. In the phase basis, a delay line interferometer (DLI) with a delay matching the distance between two successive pulses (or half a symbol) is used to measure the "visibility" $V$ of the photon states. The visibility depends on the phase between two successive pulses. An attack would decrease the coherence between those pulses, thus decreasing $V$. Assuming a collective two-pulse attack, as described in [12], the so-called secret key fraction $r$ in the infinite-key-length limit is

$$r = 1 - h(Q) - Q - (1-Q)h(1+\Delta),$$

with

$$\Delta = (2V-1)e^{-\mu} - 2\sqrt{V(1-V)}\sqrt{1-e^{-2\mu}},$$

and

$$h(p) = -p\,\log_2(p) - (1-p)\,\log_2(1-p),$$

where $Q$ is the quantum bit error rate (QBER) in the time basis, and $\mu$ is the mean photon number of a pulse. The secret key rate $S$ can then be calculated from $r$ and the sifted key rate $R$ to be $S = rR$.

In the presented work, the above-described COW protocol was modified such that multiple QKD signals can be transmitted simultaneously employing WDM. The implemented design for a simplified WDM QKD system is shown in Figure 1. On Alice's side, multiple continuous wave (CW) distributed feedback (DFB) lasers – as commonly used for wavelength-division multiplexing in classical communication – were used to create multiple QKD signals with wavelengths in the C-band around 1550 nm. The wavelengths were chosen approximately according to the WDM wavelength grid. The intensity of the CW signals was modulated with a Mach-Zehnder modulator (MZM), creating the desired pulses. For the presented proof-of-principle implementation, all QKD signals were modulated with the same MZM and thus with the same repeating pattern of sent QKD symbols. In a real-world implementation, the sending pattern of each channel should be independent. This would require $n$ MZMs, which are all modulated independently with different random symbol patterns. However, this is beyond the scope of the presented proof-of-principle experiment. The modulated signal is then multiplexed and sent to Bob.

To decrease the complexity and cost of the QKD system, a simplified receiver design was developed and implemented, where the different QKD channels share components, as will be elaborated in the following. First, the incoming signal is split by a 10-dB fiber-coupler to be evaluated in the time or phase basis, with the largest portion of the signal being routed to the time base measurement.

A straightforward approach for multiplexing multiple QKD channels based on time-phase coding would be to use one DLI on the receiver side per quantum communication channel. Thus, multiplexing multiple quantum channels would result in a huge number of DLIs. One approach to reducing the number of DLIs would be to apply the so-called "colorless" interferometric technique [14], where only one DLI on the receiving side for all quantum channels is needed. A drawback of this technique is a wavelength dependency on the DLI performance. Thus, this approach requires precise tuning of the phase of each QKD channel with respect to the DLI, resulting in the need for at least $n - 1$ additional phase modulators on the sender side.

Here, we demonstrate that a single DLI on the receiver side is sufficient for multiple quantum communication channels with a different technique. Instead of using phase modulators for phase stabilization, wavelength tuning of the deployed commercial off-the-shelf DFB Lasers was used. The tuning resolution, stability, and range of the DFB lasers allow for maintaining high interference contrasts that enable reliable QKD transmission.

With this technique, the DLI can be implemented in the signal path before de-multiplexing, such that only one DLI is necessary for evaluating the phase basis. For the time basis, the signal is directly multiplexed and forwarded to the SNSPDs. The disadvantage of the need to employ two de-multiplexers is largely outweighed by the economization of DLIs.

## 2.2 Single-photon detector system

The detection of single photons is still one of the major challenges of today's quantum technologies and brought up different approaches like the PMT, SPAD, TES and the SNSPD. If highest performance is required, SNSPDs offer superior performance that is beneficial for demanding applications such as long-range QKD. On the downside, they are arguably cost-efficient and scalable and thus are rarely used in commercial implementations [15–17].

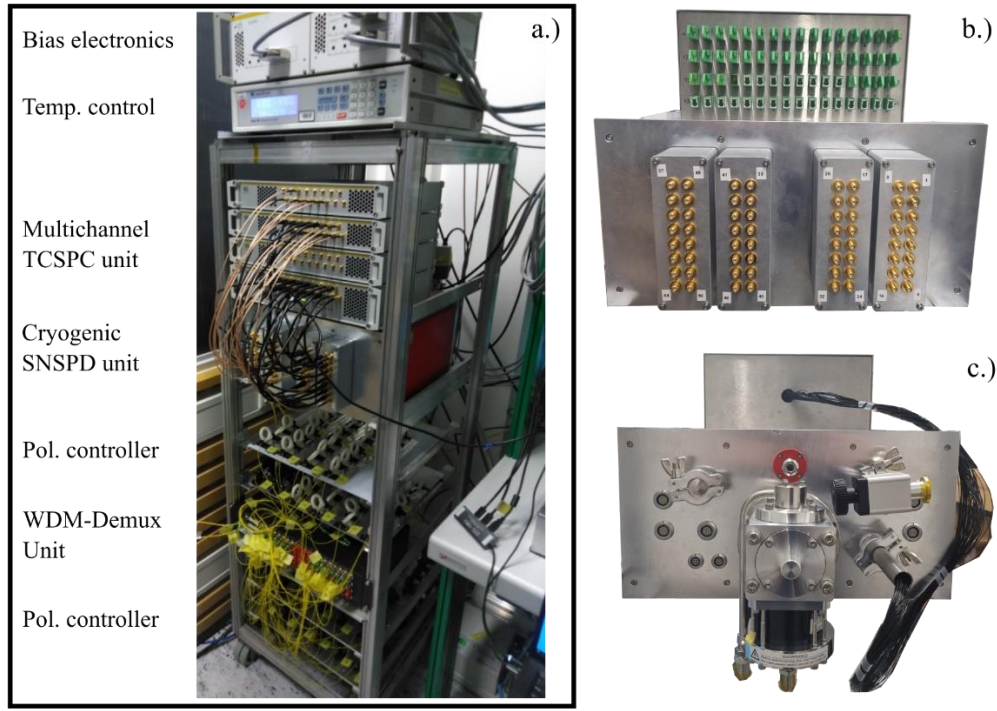

Figure 2: **(a)** 64 channel QKD receiver setup including bias electronics for amplifiers and SNSPDs, temperature control system, multichannel TCSPC unit, cryogenic SPSND unit and the WDM unit for de-multiplexing along with Fiber polarization controllers. **(b)** Front view of the SNSPD unit with LSH/APC connectors for the optical input and SMP connectors for the electrical signal output. **(b)** Back view of the SNSPD unit with feedthroughs for the optical fiber array, access openings for the vacuum pump system and electrical connectors to bias SNSPD and amplifier circuitry.

Addressing the previously mentioned technological challenges, we developed a compact detector system based on waveguide integrated superconducting nanowire single-photon detectors (WI-SNSPDs), allowing for scalability of up to 64 channels, which is depicted in Figure 2 b.) and c). The integration to waveguides enables for pre-characterization of an effective higher number of nanowires which were fabricated from a 4.4 nm thick film of superconducting NbTiN using electron beam lithography. In a subsequent process, the waveguides were routed such that the photonic input ports are connected to preselected nanowires of similar characteristics to avoid varying detection properties and maintain a high yield. The waveguide input ports for the fiber-to-chip interface, which are arranged in a 2D grid, are equipped with 3D polymer couplers made by direct laser writing [18,19] and positioned below the facet of a 2D-fiber array later on (compare Figure 3) and thus allowing for broadband coupling of light resulting in a high SDE over a wide wavelength regime [20]. The packaging order begins with electrically interfacing the detector chip via wire bonds to a printed circuit board and then attaching the fiber array above the detector chip. In the last step, the resulting module is mounted inside a rack-sized cryostat that is cooled by an air-cooled closed-cycle He-4 system. The small form factor system includes a custom hybrid cryogenic-room-temperature pre-amplification chain that allows for biasing and reading out the different channels individually and continuously operating the detector module without latching. In an initial study and before transportation, the system achieved an average SDE of 40 % for up to 37 channels, ranging from 20 % to 60 % for the individual detectors. The detectors showed count rates of up to 20 MHz with timing precisions below 120 ps and dark count rates (DCR) below 150 Hz at 3.6 K base temperature [21]. A comprehensive analysis of the packaging method, as well as an in-depth study of the achieved performance characteristics of the detector system, can be found in [22]. After evaluation of the system characteristics, it was shipped and integrated into the QKD experimental setup and built up as depicted in Figure 2 a.).

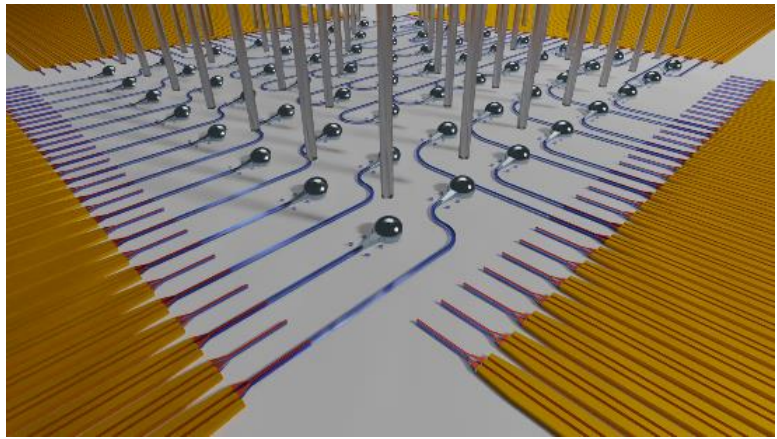

Figure 3: Schematic of a 2D-fiber array above a photonic circuit connected to WI-SNSPDs as realized in the 64-channel detector system. The fiber cores (grey, transparent) are aligned above polymer coupling structures (glossy) that are connected via waveguides (blue) to SNSPDs (red) and electrically interfaced by electrodes (gold). Note that a different scale was chosen to illustrate the different details.

The system is designed to operate up to 64 detection channels in parallel, but during the integration into the QKD experiment setup, the configuration was reduced to 32 channels due to issues with the cooling system. The cause for the monitored instabilities could be identified as the changed laboratory operation environment that reduced the cooling power and resulted in an increased base temperature of 3.8 K with a constant drift towards higher temperatures. Due to the observed drift of the cooling system, the number of detection channels, unfortunately, had to be further reduced to 22 to maintain detection performance within the required specification.

*2.3 Time Tagging System with external FPGA interface*

For the detector readout with high throughput time tagging, we developed a new scalable system with up to 64 independent but synchronized channels. The original concept of time tagging TCSPC builds on a modification of classic TCSPC electronics permitting just start-stop measurements and immediate histograms. For the first time-tagging instruments, the start-stop timing circuitry was used as previously, providing the required picosecond time resolution for TCSPC. In order to maintain the full information of the temporal patterns of photon arrivals, the events were no longer stored as histograms but as separate records. In addition, a coarser timing was performed on each photon event with respect to the start of the experiment [23]. This is referred to as time-tagged time-resolved (TTTR) data collection or more generically today, just "time tagging". The TTTR concept avoids both, redundancy in the data stream and loss of information. As a result, virtually all algorithms and methods for the analysis of photon dynamics over different time scales can be implemented. There is a vast range of methods and applications building on this concept in the area of life sciences [24]. In the context of quantum optics, the same high-resolution global arrival time tagging of photon detections is similarly valuable. For instance, the generation of suitable states of light via single-photon sources is one of the most important tasks at hand [25–27]. Here the second-order correlation measurement plays an important role – the depth of the dip for zero delay is a direct indicator of how well such a single photon source performs. For a single-photon state, $g^{(2)}(0)$ should be zero, and recent solid-state sources come very close to this value [25,28]. These correlation measurements also lie at the heart of experiments in fundamental quantum mechanics [29], e.g., the ability to extract even higher order correlations from photon arrival times is routinely exploited. Ideally, such correlation measurements are done on independent timing channels such that dead time effects can be eliminated by cross-correlating the detector signals [30]. In quantum communication, the signal-to-noise ratio can be improved drastically by identifying and eliminating photons from background processes in the time tag analysis [31]. Similarly, time tagging is a routine tool in research on QKD protocols and systems [32,33]. Our new implementation of a time tagging system for the task at hand conceptually builds on earlier work where the number of channels was still limited to 16 and the digital time resolution was only 80 ps [11]. In order to meet the goals for the QKD system described here, it was necessary to extend the number of channels to 64, which was achieved by a modular design employing multiple FPGA modules, each handling a row of 8 timing channels. The modules are fully synchronized so that the overall system behaves like a monolithic solution, delivering a single data stream with the events from all channels arriving in their correct temporal order. This has great benefits in the real-time processing of the data. The FPGA-based TDCs were significantly improved to achieve a digital resolution of 5 ps, maintaining the very short dead time of only 650 ps.

In addition to the 64 regular timing channels, the system provides a common synchronization channel. This channel serves as a timing reference in classic histogramming measurements and the so-called T3 mode, which is explained further below. The sync channel provides the same resolution and dead time as the regular timing channels and may, if no common sync is required, be used as an additional detector channel. In order to support high sync rates, e.g., form a fast laser, the sync channel provides a programmable divider permitting sync rates as high as 1.2 GHz.

Furthermore, in addition to the high-resolution timing inputs, the TCSPC electronics provide four inputs for TTL signals that are captured at a lower resolution and get inserted in the data stream exactly like regular timing events. These low-resolution signals can serve as markers for different schemes of secondary synchronization, e.g., representing spatial information of scanning devices or control events in a QKD system.

As in our previous systems, a central crystal clock ensures that all timing inputs have a common time base. Optionally, the clock may also be fed in as an industry standard 10 MHz signal from an external source such as an atomic clock or a GPS receiver. The same standard clock signal is also available as an output so that other devices can be daisy-chained. Similarly, the new time tagger provides a White Rabbit interface, allowing long-distance remote synchronization over Gbit Ethernet fiber links that may also carry regular TCP/IP traffic. It was previously shown that the White Rabbit protocol could achieve a synchronization precision of a few tens of picoseconds [34]. Recently we also demonstrated that our implementation is interoperable with commercial White Rabbit Ethernet switches and that the same level of precision can be achieved across switched networks of various topologies [35]. Both the synchronization via 10 MHz from GPS and that via White Rabbit are of great value in QKD scenarios. While GPS has the benefit of wireless transmission, White Rabbit has the benefit of substantially higher synchronization precision and the convenient simultaneous transport of data. With the latter, it should even be possible to employ wavelength multiplexing so that the secret QKD channel could use the same fiber as the public TCP/IP channel.

The data acquisition schemes of the new time tagging system follow proven concepts [11,30,36]. Apart from classic histogramming, the system provides two time-tagging modes, called T2 and T3 modes, which differ from each other by their handling of sync events. The T2 mode registers signal inputs equally for all input ports, whether the signal comes from a connected photon detector or just contains a sync signal. When the sync input is used for a detector, the divider can be bypassed. For all registered events, which in T2 mode are recorded and handled equally, a 32-bit record containing the information about the channel number and the arrival time after the start of the measurement with a resolution of 5 ps is generated. To achieve continuous operation, even when the number of bits reserved for the time tags is exceeded, an overflow record is inserted into the data stream, and continuous processing of the data is thereby possible for any desired period.

For each channel, the timing records from the TDCs are stored in separate front end FiFo buffers, which are fast enough to accept records at the maximum speed of the TDCs of 1.53 GHz. Only if the space of up to 2048 records in the FiFo runs out events are dropped, which will be notified to the host computer by a corresponding flag. An additional, considerably larger FiFo buffer, in which the temporally sorted T2 records from all channels are queued, can store up to 268,435,456 event records and is continuously read via a USB 3.1 Gen 1 interface by the host PC. The FiFo input can cover bursts of a much higher rate and allows to maintain the data integrity if the average data rate to the host is not exceeded for a long time. In the case of such a FiFo overrun, the measurement is aborted. Nevertheless, sustained average rates of 88 Mcps (shared by all input channels) were achieved with a recent host system (Windows 10 Intel Core i7 7740X @4.3 GHz).

The T3 mode is conceptually similar to the TCSPC histogramming mode. It is typically used when the sync channel is needed for an excitation source of a high repetition rate. Frequently used mode-locked laser systems with tens of MHz sync rate would overrun the FiFo buffer very quickly if processed in T2 mode. In T3 mode, only the timing records but not the actual sync events are forwarded. The timing records in T3 mode contain a coarse time tag derived from the counted sync pulses and a high-resolution time difference of the start-stop measurement between sync and photon event. The data records are held in a 32-bit format, also including information on the channel number.

If the sync count exceeds what the limited number of bits can hold, an overflow record is inserted, allowing unlimited measurement times as in T2 mode. The covered time span of the high-resolution start-stop measurements depends on the chosen resolution R for the time bins of 5 ps to 41.9 μs. With a width of 15 bits for the start-stop record, the resulting time span covers a period of 32768 × R, which corresponds to 163.84 ns at the highest resolution and 1372 ms at the lowest resolution. Events of even higher time delay than this time span are not recognized by the system, just as for the classical histogramming mode that is similarly limited by the number of time bins. The T3 mode records are read by the host system from the same FiFo as for the T2 mode, achieving burst buffering and similar maximum average count rates for every individual channel. In contrast to T2 mode, the bus is not loaded with transfers of sync events.

The USB interface provides good performance and flexibility for many use cases. However, when targeting very high count rates in setups using all 64 channels, as in the QKD application here, its bandwidth becomes limiting. Furthermore, the host computer must keep up with the generated data, which can be a challenge when running complex real-time data processing algorithms. In order to address this issue, an external FPGA Interface (EFI) was developed. It uses two or more high-speed serial links (6 Gbits/s each) to transfer data to an external FPGA, where fast parallel data processing and I/O can be performed with virtually unlimited flexibility. Custom logic can be written in VHDL, Verilog or any other language supported by Xilinx Vivado, it only needs to be connected to a set of high-level data stream interfaces. The EFI can deliver raw timing data or pre-processed T2 or T3 mode data. Through a loopback interface, custom data generated by the external FPGA can even be transmitted back to the time tagger unit and retrieved through the USB interface. Deep customization of the gateware IP is supported, as all required sources for the external FPGA are made openly available.

The most convenient mode of operation of the EFI is equivalent to T2 or T3 mode over USB, only with higher bandwidth. It uses two physical serial links aggregated as one 12 Gbits/s logical link over which data is streamed in T2 or T3 format, with records of all (up to 64) channels temporally ordered. To achieve this temporal ordering, the data of all timing modules is passed through the main unit for sorting and aggregation.

The fastest mode of operation of the EFI is the T2 direct mode (T2DM), where each of the four time tagger modules (16 channels each) delivers its data directly via two serial links (6 Gbits/s each) to the external FPGA. Using T2DM, it is possible to process more than 1.6 G events/s using all 64+1 channels. It also reduces the latency between measurement and availability in the external FPGA by 80 % compared to the regular T2/T3 streams. The downside is that in this case, the data is not temporally ordered and must be sorted in the external FPGA if their temporal relation across the individual timing modules matters. The T2DM streams are based on the semantics of the T2 mode.

| Mode | Throughput | | Latency [µs] |
|---|---|---|---|
| T2 | 200.000.000 event/s | | 4.5 to 5.0 |
| T3 | 200.000.000 event/s | | 4.5 to 5.0 |
| T2DM | 78.000.000 event/s | for the sync + | 1.7 to 1.8 |
| | 200.000.000 event/s | shared among each row of 8 inputs | 0.8 to 1.2 |

Table 1: Performance Figures of the different EFI modes

## 2.4 Real-time QKD evaluation on FPGA

To evaluate the QKD transmission, a test pattern with a random but repeating pattern was modulated with the MZM. The transmitted QKD test pattern consists of 96 symbols, where the symbols "0" and "1" occur 43 times each and the Decoy symbols 10 times to roughly match the 10-dB fiber-coupler used as the receiver's basis choice. The symbols are randomly distributed within these 96 symbols. The evaluation was performed in real-time on the previously mentioned external FPGA (here, a Digilent Genesys 2 Kintex®-7 FPGA development board). The evaluation was implemented in FPGA logic (not in a softcore). The data was accumulated over a one-second interval. The accumulated data was then evaluated during the next time interval, in parallel to accumulating data for the next interval. Note that the evaluation took only around 20 msec of processing time, which allows for additional or more advanced processing if needed.

The evaluation was performed by comparing either the known sending pattern (for the time basis) or the consequential interference pattern (for the phase basis) with the pattern received by Bob. For that purpose, the incoming signal was measured in T3 mode. The clock signal was transmitted by the sender and clicked with every repetition of the sending pattern.

The detection signals were sorted by the time slot to build a histogram, which was then compared to the sending or interference pattern, respectively. By counting and processing those events that agreed or disagreed with the sending- or interference pattern, the sifted key rate, Quantum Bit Error Rate (QBER) and visibility were calculated. Since the relative shift between Alice's sent pattern and Bob's received pattern varies with the transmission distance, the evaluation was repeated for all possible shifts. Then, the smallest QBER and highest visibility correspond to the correct shift and thus yield the correct results.

Based on these measurement results, the secret key rate was then calculated offline. Note that in order to improve the QKD performance further, detections within the first and last 40 ps of every 400 ps long slot were discarded and ignored in the evaluation. This served as a filter to only process the most unambiguous photon clicks, thus decreasing QBER and increasing visibility at the cost of only a small decline in the sifted key rate. As a result, both the secret key rate and the range of the QKD transmission increased.

## 3. Experimental results

### 3.1 Time tagger performance tests

The new time tagging system was tested in various scenarios to verify functionality and timing accuracy. Even though the digital TDC resolution is fixed at 5 ps, a timing uncertainty (jitter) due to noise is always present. This is the case even for digital signals because of the finite slope of the signal transitions. In order to test this quantity for the prototype design, the following test was performed. A test generator (CG635, Stanford Research Systems) delivered pulses of 10 MHz repetition rate with rise/fall times of 370 ps (10 to 90 %). The steep transitions ensure that the time measurement results are only insignificantly influenced by noise, as shown by earlier measurements on TDCs with substantially higher resolution and precision [37]. This signal was fanned out through a reflection-free splitter so that five identical signals were obtained. These were fed to the sync input and four input channels of the device under test. The device was then operated in histogramming mode, where the time differences between sync and the respective input channel are recorded. The result is shown in Figure 4.

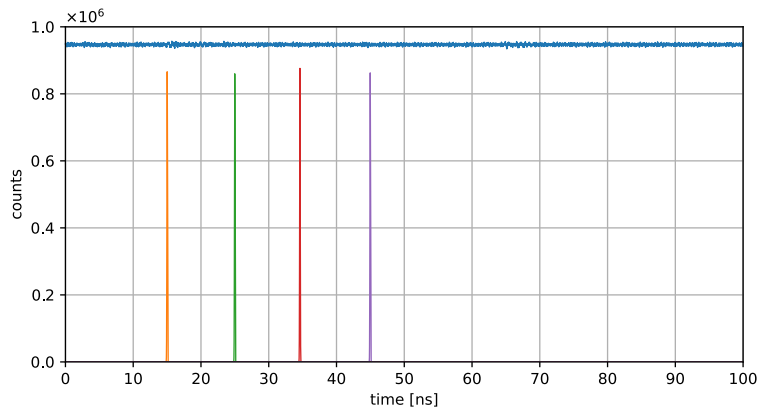

Figure 4: Constant delay time difference histograms for four exemplary channels versus sync (sharp peaks at about 15, 25, 35 and 45 ns) and result of an exemplary DNL measurement (flat line at the top) of the new time tagger.

Each peak at 15, 25, 35, and 45 ns represents the histogram obtained for one channel. Using the software adjustable offset of each channel, the 10 ns spacings between the peaks were set arbitrarily for a clear view. Numerical analysis of the distributions shows that the r.m.s. timing jitter is typically 43 ps. It should be noted that this is the overall error, including the sync channel and the respective detector channel. The single channel measurement jitter would correspond to 43 ps / $\sqrt{2}$, which is about 30 ps.

The next important key characteristic is the differential nonlinearity (DNL), which is a measure of the systematic error of the bin widths of the TDCs. For a quantification measurement of the DNL, the correlation of two different pulse generators, one as sync reference with a frequency of 10 Mhz (Staford Research System, CG635), and one as independent signal source with a 1 MHz pulse train and period dithering of 8% (Dallas Semiconductor, DS1090), were used. The expected start-stop histogram should therefore show uniform counts for the different time bins, only deviated by the statistical or systematical (DNL) error.

To keep the statistical error at a minimum, the experiment was run until a total number of about $10^6$ counts was reached (compare Figure 4). Analysis of the measurement attests a r.m.s. deviation below 2.5 % peak-to-peak, averaging at 0.3 %. The achieved performance of the TDCs is by orders of magnitude lower than other common designs for TDCs that often show DNL errors of up to 100 %.

### 3.2 QDK transmission results

For an investigation of the system performance in a QKD application, the multichannel SNSPD unit and the multichannel time tagger unit were combined with the COW-QKD setup. In between Alice's and Bob's setup, an optical attenuator was used to experimentally simulate the transmission losses of common glass fibers to evaluate the long-range QKD capabilities. The basic level for the attenuation was set to a value according to the desired average number of photons per pulse of $\mu \leq 0.1$ for all channels to ensure a single-photon level intensity. For a back-to-back transmission, no further attenuation was added and sifted key rate, QBER, visibility and secret key rate were measured and tracked for a period of 35 minutes (Figure 5). For the most part, the QKD system shows consistent performance with secret key rates between 0.6 to 3.2 Mbit\s. As can be seen in Figure 5, some minor impairment is caused by a slow drift of the interferometer phase, which is not actively stabilized. However, since this is only one common parameter for all QKD channels, it could be stabilized automatically very easily. The differences in the achieved key rates are mainly caused by efficiency differences of the SNSPD. For two QKD channels, one of the SNSPDs of the two channels suffered from the increasing temperature of the cryogenic system inhibiting reliable QKD transmission. Therefore, only the results of 9 QKD channels were evaluated.

To evaluate the long-range capabilities of the QKD system, additional dampening was added to simulate the losses induced by optical fibers. The system achieves secret key rates starting from 13.2 Mbit\s for all channels without additional loss and can maintain operation for attenuation of up to 26.6 dB with a secret key rate of 3.8 Mbit\s (Figure 6) and corresponding to the losses of a 130-km-long standard single-mode fiber.

The different QKD channels should, to some extent, show very similar long-range QKD capabilities since the dark count rate, the main limiting factor for this performance aspect, was measured between 100 to 150 Hz for all detection channels before any QKD measurement was carried out. Due to the increasing temperature, the DCR began to deteriorate for some channels, increasing the window of DCRs to 0.1-10 kHz. For the highest attenuation of 26.6 dB, only one QKD channel remained operational.

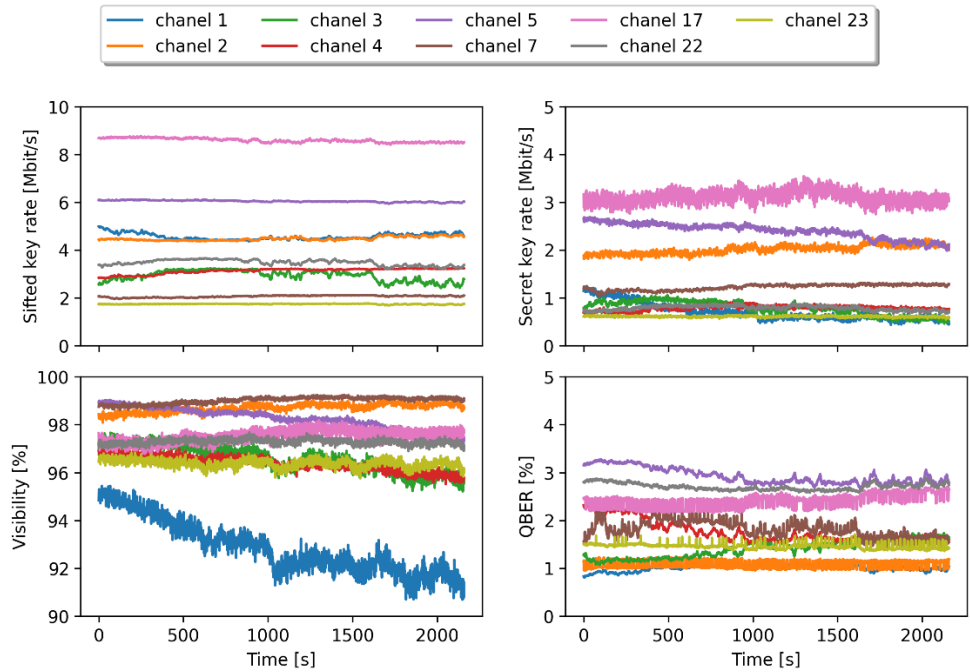

Figure 5: QKD transmission results of a back-to-back measurement (average number of photons per pulse set to $\mu \leq 0.1$) for 9 QKD channels. The sifted key rate (top left), secret key rate (top right), visibility (bottom left) and QBER (bottom right) are shown for each wavelength channel for a 35-minute continuous transmission.

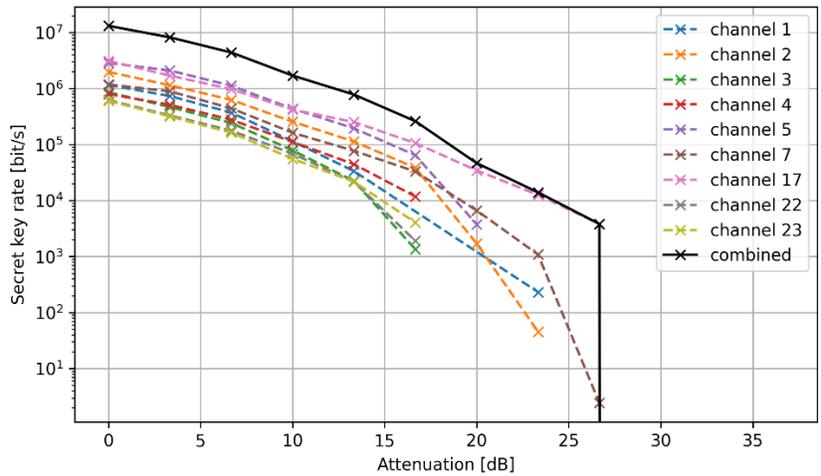

Figure 6: The secret key rates were averaged for over 1 minute for each attenuation in the detection channel (dashed colored lines). The total secret key rate of all channels (solid black line) declines from 13.2 Mbit\s without additional attenuation to 3.82 kbit\s at 26.6 dB attenuation.

## 4. Conclusion

The focus of this work was to implement scalable solutions for the different parts of a COW-QKD on a high number of channels. A QKD setup was built using a COW protocol utilizing

WDM to demonstrate QKD on up to 40 wavelength channels in parallel. For the detection of photons, a mobile air-cooled cryogenic single photon detection system offering up to 64 channels was developed. In a first configuration, 37 single photon counting channels satisfy the requirements for ultrafast COW-QKD. Secondly, to sense and time tag the signals of up to 64 detectors, a new scalable TCSPC system with 64 independent but synchronized channels was developed and thoroughly characterized. The TCSPC unit can register up to 1.6 Giga event/s with a timing jitter of 30 ps. To evaluate the extremely high data rates at once, the system features a high-bandwidth serial link interface that was later used to process the time bins of a QKD transmission with an FPGA evaluation board. The final QKD experiment was performed by combining the novel multichannel approaches in a single rack structure. Due to unexpected thermal limits in the cooling system of the detector unit, a successful QKD transmission was carried out on 9 wavelength channels utilizing 18 single photon counting channels with a total maximum back-to-back secret key rate of 13.2 Mbit\s or 3.82 kbit\s for 26.6 dB attenuation.

**Funding.** The authors acknowledge the financial support by the Federal Ministry of Education and Research of Germany in the framework of QuPAD (BMBF 13N14953).

**Disclosure.** The authors declare no conflicts of interest.

**Data availability.** The data that support the plots are available from the corresponding authors upon reasonable request.